\documentclass[aps, reprint, superscriptaddress, prb, preprintnumbers, amsmath, amssymb, letterpaper, floatfix, showpacs]{revtex4-2}
\usepackage{amsmath, amssymb}
\usepackage{graphicx}
\usepackage{times}
\usepackage{dcolumn}
\usepackage{multirow}
\usepackage[colorlinks, linkcolor=blue, anchorcolor=blue, urlcolor=blue, citecolor=blue, pdfborder=001]{hyperref}

\newcommand{\IrGa}{Ir$_{2}$Ga$_{9}$}
\newcommand{\musr}{$\mu$SR}
\newcommand{\Tc}{$T_{\rm{c}}$}

\begin{document}


\title{Type-I/Type-II superconductivity in noncentrosymmetric compound Ir$_{\mathbf{2}}$Ga$_{\mathbf{9}}$}

\author{J. C. Jiao}
\author{K. W. Chen}
\affiliation{State Key Laboratory of Surface Physics, Department of Physics, Fudan University, Shanghai 200438, People's Republic of China}
\author{O. O. Bernal}
\affiliation{Department of Physics and Astronomy, California State University, Los Angeles, California 90032, USA}
\author{P. -C. Ho}
\affiliation{Department of Physics, California State University, Fresno, California 93740,USA}
\author{Lei Shu}
\altaffiliation[Corresponding Author: ]{leishu@fudan.edu.cn}
\affiliation{State Key Laboratory of Surface Physics, Department of Physics, Fudan University, Shanghai 200438, People's Republic of China}
\affiliation{Shanghai Research Center for Quantum Sciences, Shanghai 201315, People`s Republic of China}

\date{\today}

\begin{abstract}
We have performed magnetization, specific heat, and muon spin relaxation (\musr) measurements on single crystals of the noncentrosymmetric superconductor \IrGa. The isothermal magnetization measurements show that there is a crossover from Type-I to Type-II superconductivity with decreasing temperature. Potential multi-band superconductivity of \IrGa~is observed in the specific heat data. \musr~measurement is performed to map the phase diagram of \IrGa, and both Type-I and Type-II superconductivity characteristics are obtained. Most importantly, a more unique region with the coexistence of Type-I and Type-II \musr~signals is observed. In addition, time reversal symmetry is found to be preserved in \IrGa~by zero field \musr~measurement. 
\end{abstract}


\maketitle

\section{Introduction}

Superconductivity in materials with noncentrosymmetric (NCS) structure has been widely studied due to the theoretically proposed unconventional pairing state~\cite{Naoto2009, Smidman2017, Shang2022}. In NCS superconductors, an electronic antisymmetric spin-orbit coupling (ASOC) is allowed to exist due to the absence of inversion symmetry. A sufficiently large ASOC has profound consequences on the superconducting state, and may lead to a mixed singlet-triplet nature in the superconducting order parameter~\cite{Bauer2012}. As a result, it can give rise to a range of novel phenomena, including unconventional gap symmetry with nodes~\cite{Hirata2007, Samokhin2008}, topological superconductivity~\cite{Hyunsoo2018, Zhu2022}, and potential time reversal symmetry breaking (TRSB)~\cite{Hillier2009, Barker2015}. However, the degree of correlation between these novel phenomena and NCS structure needs further research to confirm, especially for TRSB.

Meanwhile, superconductors are generally classified as Type-I and Type-II according to the Ginzburg and Landau (GL) paradigm. The Ginzburg-Landau parameter $\kappa=\frac{1}{\sqrt{2}}$ is the boundary of Type-I and Type-II classification~\cite{Tinkham2004}. For $\kappa<\frac{1}{\sqrt{2}}$, superconductors are considered to be Type-I, exhibiting Meissner, intermediate, and normal states in the phase diagram. For type-II superconductors in which $\kappa>\frac{1}{\sqrt{2}}$ in external fields, quantized magnetic flux enters into the sample, forming Abrikosov vortices. A stable vortex lattice can be formed due to the repulsive interactions between the flux vortices, and such a state is known as a mixed state~\cite{Marlyse1996}. However, when $\kappa$ is close to $\frac{1}{\sqrt{2}}$, the microscopic correction to the GL theory proposes an attractive interaction between flux vortices~\cite{Eilenberger1969, Kramer1971}, leading to the emergence of  a new state called intermediate-mixed state (IMS)~\cite{Jacobs1973, Christen1980}. In that case, superconductors may exhibit both Type-I and Type-II characteristics, which is generally described as Type-I/Type-II behavior. Superconductors showing Type-I/Type-II behavior, such as MgB$_2$~\cite{Kortus2001} and ZrB$_{12}$~\cite{Lortz2005}, have been widely studied due to their various unconventional superconducting behavior. In addition, the Type-I/Type-II behavior has also been found in some NCS superconductors, such as NbGe$_2$~\cite{Lv2020} and LaRhSi$_3$~\cite{Kimura2016}. However, there is a lack of in-depth research dealing with the correlation between the Type-I/Type-II behavior of these superconductors and their NCS structures.

Superconductivity with \Tc~= 2.25 K in \IrGa~was first observed in 2007~\cite{Shibayama2007}. \IrGa~crystallizes in a distorted Co$_2$Al$_9$-type structure, which lacks spatial inversion symmetry. However, there is still some controversy about its superconducting properties. \IrGa~polycrystal sample was suggested to be a Type-II superconductor, since the estimated GL parameter $\kappa$ is close to $\frac{1}{\sqrt{2}}$ using the ratio of its upper and lower critical fields~\cite{Shibayama2007}. However, the value of $\kappa$ obtained from specific heat as a function of magnetic field from \IrGa~the single crystal sample suggested that \IrGa~is a Type-I superconductor~\cite{Wakui2009}. 

To identify the nature of superconductivity of \IrGa, a detailed study was conducted on the physical properties of  a single crystal of \IrGa. Based on the susceptibility measurements, \IrGa~shows Type-I superconducting behavior below \Tc,  but some Type-I/Type-II behavior is revealed at low temperatures. Potential multi-band superconductivity evidence is suggested by the low-temperature specific heat data. In addition, the muon spin relaxation (\musr)~technique measures the internal magnetic field distribution, and has been widely used to map the phase diagram and study the microscopic properties of both Type-I and Type-II superconductors~\cite{Sonier2000, Beare2019, Karl2019, Leng2019, Biswas2020}. \musr~measurements on single crystals of \IrGa~have been carried out to measure the inhomogeneous field distributions, and map the different superconducting states in the $H$-$T$ phase diagram. Our studies reveal that \IrGa~exhibits the coexistence of \musr~signals typical of both Type-I and Type-II superconductors. Meanwhile, thanks to its high sensitivity of magnetic field, \musr~has a great advantage in exploring the potential TRSB superconductivity in NCS superconductors. Our measurements suggest that the time reversal symmetry is preserved in \IrGa.

\section{Experiments}

\IrGa~single crystals were grown by the Ga flux method as previously reported~\cite{Wakui2009}. The typical size of one single crystal piece is about 2~mm$\times$1~mm$\times$1~mm. The chemical composition and crystal structure are checked by the electron probe microanalysis (EPMA) and Laue measurements. Susceptibility measurements were carried out using a commercial vibrating sample magnetometer (Quantum Design) from $T$ = 1.8~K to 3.0~K. The isothermal magnetization 4$\pi$$M(H)$ was measured in the temperature range of 1.8~K to 2.3~K. There is not a specific crystal orientation in the measurement. The specific heat of \IrGa~from $T$ = 0.45~K to 4.0~K was measured in a Quantum Design physical property measurement system (PPMS) using the relaxation method.

Zero-field (ZF) and transverse-field (TF) \musr~experiments were carried out at the M15 beam line, TRIUMF, Vancouver, Canada. About 50 pieces of \IrGa~single crystal samples were mounted on a silver sample holder with random orientations. ZF-\musr~was performed above and below \Tc~down to $T$ = 0.05~K to study whether there is spontaneous small magnetic field in the superconducting state due to the TRSB superconductivity. In TF-\musr~experiments, several temperature scan ($T$-scan) measurements were carried out at different applied magnetic fields to map the phase diagram of \IrGa. The magnetic field was applied after the sample was first cooled down to the lowest temperature in zero field. The \musr~data were analyzed by using the MUSRFIT software package. 

\section{Results}

\subsection{Magnetization}

\begin{figure}[ht]
	\begin{center}
		\includegraphics[clip=,width=8.5cm]{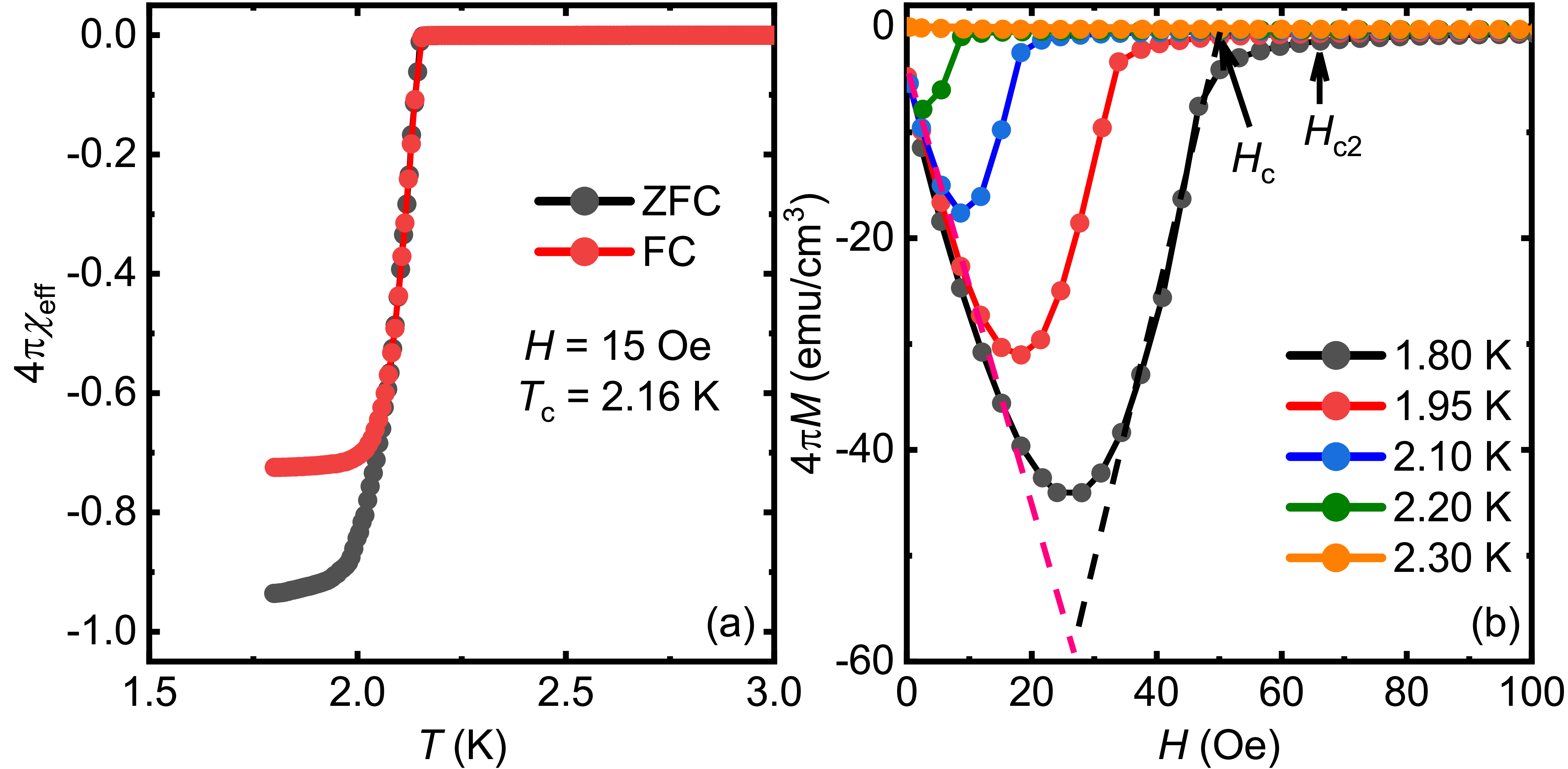}
		\caption{(a) Temperature dependence of the dc susceptibility $\chi(T)$ of \IrGa~in the ZFC and FC modes showing superconductivity at \Tc~= 2.16 K in an external field $H$ = 15 Oe. (b)  Magnetization $M(H)$ curves at various temperatures. }
		\label{fig: Mag}
	\end{center}
\end{figure}

The dc magnetic susceptibility measured under a magnetic field $H$ = 15 Oe in both zero-field-cooling (ZFC) and field-cooling (FC) modes are shown in Fig.~\ref{fig: Mag}(a). Superconductivity is observed below the onset point of the diamagnetic signal, \Tc~(15 Oe)~= 2.16 K. Due to the relatively small critical field of \IrGa, even at a small external field of 15 Oe, \Tc~ is significantly suppressed compared to the zero field situation, i.e., \Tc(0) = 2.25~K determined from the specific heat measurement. The superconducting volume fraction estimated from the ZFC data is close to 100\%, indicating high sample quality. In general, due to the magnetic flux pinning effect in FC modes, there will be a significant deviation in the FC and ZFC magnetic susceptibility curves for Type-II superconductors. On the contrary, for Type-I superconductors, the ZFC and FC magnetic susceptibility curves are highly consistent. For \IrGa, the degree of deviation between ZFC and FC curves is between typical situations of Type-I and Type-II superconductors~\cite{Lin2016, Peets2019}. A similar phenomenon has been observed in several Type-I/Type-II superconductors, such as ZrB$_{12}$~\cite{Lortz2005}, indicating that \IrGa~may exhibit the Type-I/Type-II superconducting property. However, in real situation, the pinning may also be present in Type-I superconductors. And there are other effects, for instance, the so-called 'topological hysteresis'~\cite{Prozorov2005, Tejada2008}, which may lead the deviation in the FC and ZFC magnetic susceptibility curves in Type-I superconductors. Thus, more experiments are needed to detect the potential Type-I/Type-II superconductivity in \IrGa. 

The isothermal magnetization 4$\pi$$M(H)$ of \IrGa~in the temperature range 1.8-2.3 K is shown in Fig.~\ref{fig: Mag}(b). At temperatures close to \Tc, the $M(H)$ curves are consistent with that of typical type-I superconductor, as the magnetization sharply approaches zero at the critical field. At low temperatures, a deviation from the linearity of the $M(H)$ curves is observed as the magnetization approaches zero.  As shown in Fig.~\ref{fig: Mag}(b), $H_{c}$ is obtained by extending the linear part of the curve to $M = 0$, and $H_{c2}$ is defined as the point where $M$ reached 0~\cite{Lv2020}. These characteristics are consistent with the reported Type-1.5 superconductivity~\cite{Lortz2005, Kimura2016, Lv2020}. In the Type-1.5 superconductivity, the appearance of the small tail in $M(H)$ is due to the entry of magnetic flux, which leads to a mixed state. Thus, there is a crossover from Type-I to Type-II in the phase diagram, leading to the existence of the Meissner-mixed state and the so-called intermediate-mixed state. 

\subsection{Specific heat}

\begin{figure}[ht]
	\begin{center}
		\includegraphics[clip=,width=8cm]{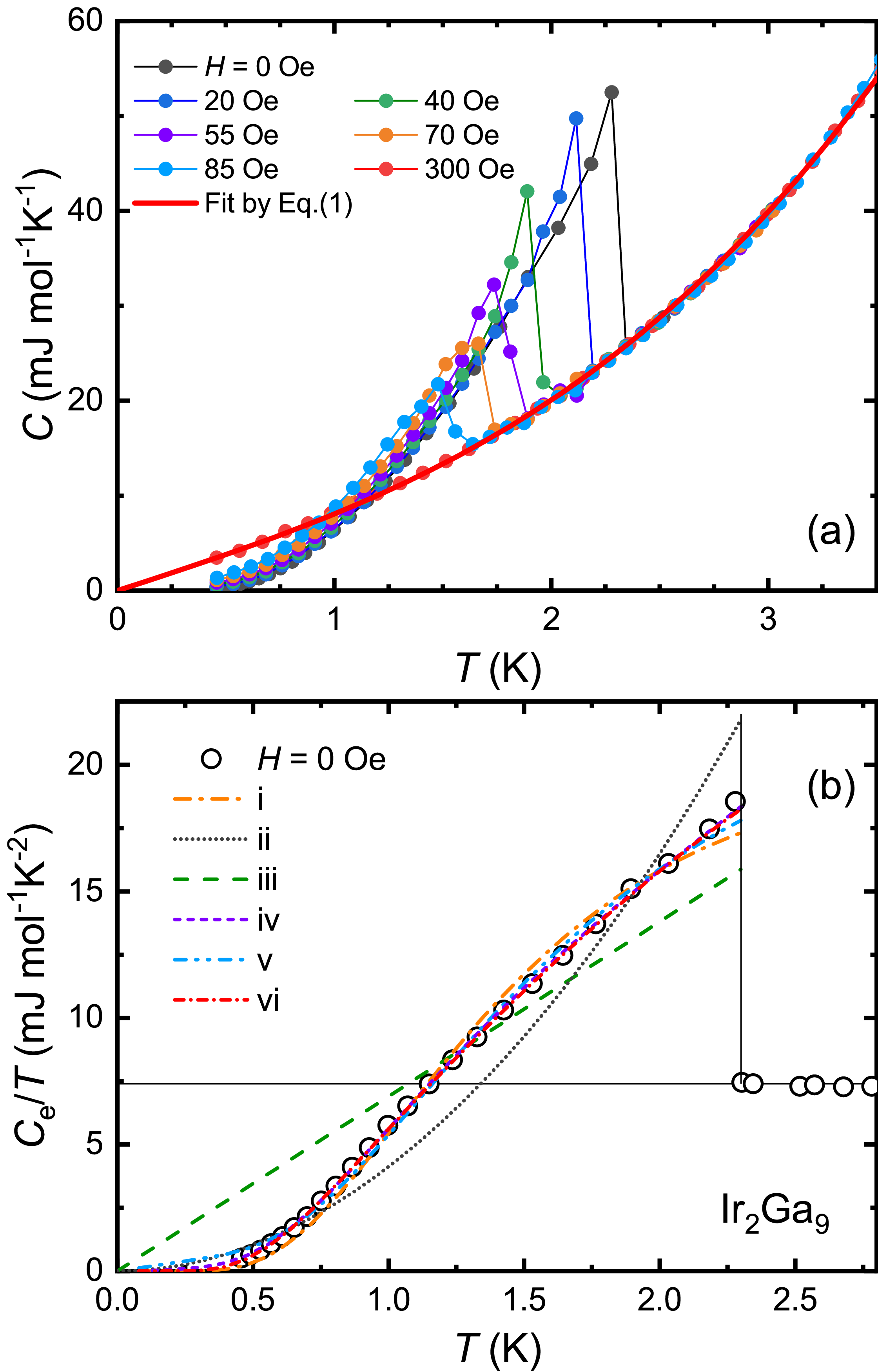}
		\caption{(a) Specific heat of \IrGa~single crystal at several applied fields from 0.45 K to 3.5 K. Eq.~(\ref{eq:HC}) is fitted to the normal state specific heat data at the applied field $H$ = 300 Oe, shown as the solid red curve. (b) Temperature dependence of the electronic specific heat of \IrGa~at zero field. The curves represent the fits using six different gap models listed in Table~\ref{tab:Cel}.}
		\label{fig: HC}
	\end{center}
\end{figure}

The temperature dependence of the specific heat $C_{\rm{p}}$ of \IrGa~measured at different magnetic fields is shown in Fig.~\ref{fig: HC}(a), showing the suppression of \Tc~ by magnetic fields. The normal-state $C_{\rm{p}}$ is well described by the expression
\begin{equation}
	 C_{\rm{p}} = \gamma T + \beta T^3
	 \label{eq:HC}
\end{equation}
 where the first and second terms correspond to the electronic and phononic contributions, respectively. By fitting the $C_{\rm{p}}$ data at 300 Oe using Eq.~(\ref{eq:HC}), we obtain $\gamma = 7.42(4)$~mJ/mol K$^2$ and $\beta = 0.652(5)$~mJ/mol K$^4$, consistent with those from a previous report~\cite{Wakui2009}. The zero-field electronic specific heat $C_{\rm{e}}$ is calculated by subtracting the phonon part from the total $C_{\rm{p}}$, and is plotted as $C_{\rm{e}}/T$ versus $T$, as shown in Fig.~\ref{fig: HC}(b). To investigate the gap symmetry of \IrGa, six models are fitted to the data of $C_{\rm{e}}$, as listed in Table~\ref{tab:Cel}. Previous work only use a single-gap BCS model to fit the $C_{\rm{e}}$ data~\cite{Shibayama2007, Wakui2009}. However, as indicated in Fig.~\ref{fig: HC}(b) and Table~\ref{tab:Cel}, single-gap BCS model may not the best model to describe our data. Instead, we use the phenomenological two-gap $\alpha$ model with a weighing factor $f$~\cite{Bouquet2001}
\begin{equation}
	\label{eq:Cel}
	C_{\rm{e}} = fC_{\rm{e,1}} + (1 - f)C_{\rm{e,2}},
\end{equation} 
where $C_{\rm{e},x}~(x = 1,2)$ is the electronic contribution from each gap. This model is often used to describe a multi-band superconductors~\cite{Fisher2003, Zhang2019}. Here $C_{\rm{e},x}\propto~e^{-\Delta/k_{B}T}$, $\propto~T^2$, and $\propto~T^3$ refers to the weakly coupled BCS gap ($s$ wave), the point-node gap, and the line-node gap, respectively.

\begin{table*} 
	\caption{\label{tab:Cel} Fitting models applied to the $T$ dependence of $C_{\rm{e}}$}
	\begin{ruledtabular}
		\begin{tabular}{cccccl}
			$\ $ & Model & $f,(1-f)$ & $\Delta/k_{B}$\Tc & Adj. $R^2$ & $\Delta_{C_{\rm{p}}}/\gamma$\Tc\\
			\colrule
			i & $s$ wave  & 1, 0 & 1.56(2) & 0.9856 & 1.34(3) \\ 
			ii & Line-node   & 1, 0 & \ & 0.9387 & 1.94(7) \\ 
			iii & Point-node & 1, 0 &\ & 0.8605 & 1.14(9) \\ 
			iv & $s$ wave + Line-node & 0.52, 0.48 & 1.36(1) & 0.9992 & 1.48(5) \\ 
			v & $s$ wave + Point-node & 0.75, 0.25 & 1.65(2) & 0.9892 & 1.42(9) \\ 
			vi &$s$ wave + $s$ wave & 0.45, 0.55 & 1.25(2), 3.4(2) & 0.9995 & 1.47(4) \\ 
		\end{tabular}
	\end{ruledtabular}
\end{table*}

As shown in Table~\ref{tab:Cel}, $C_{\rm{e}}$ of \IrGa~can be well fitted by two gap models (iv, vi), suggesting that \IrGa~may has a multi-band feature, which may be helpful to better understand the Type-I/Type-II behavior in \IrGa~theoretically. Just from the fitting parameters, besides the weakly coupled BCS gap ($\Delta/k_{B}$\Tc~$< 1.76$), \IrGa~may has another superconducting gap, which can be a strongly coupled $s$-wave or gap with line-node.  However, due to the simple form of the two-gap $\alpha$ model, the previous analysis is not sufficient to conclude that \IrGa~has multi-band features. Until there is more experiment evidence, it is more appropriate to conclude \IrGa~has a possible fully gapped $s$-wave symmetry.

\subsection{TF-\musr}

In the TF-\musr~experiment, the external field $H_{\rm{ext}}$ is applied perpendicular to the initial muon spin polarization. After each muon is injected into the sample, the muon spin precesses about the local magnetic field $B_{\rm{loc}}$ at the muon stopping site with the Larmor frequency $\omega = \gamma_{\mu}B_{\rm{loc}}$, where $\gamma_{\mu} = 2\pi \times135.53\  \rm{MHz/T}$ is the muon gyromagnetic ratio.

\begin{figure*}[ht]
	\begin{center}
		\includegraphics[clip=,width=17cm]{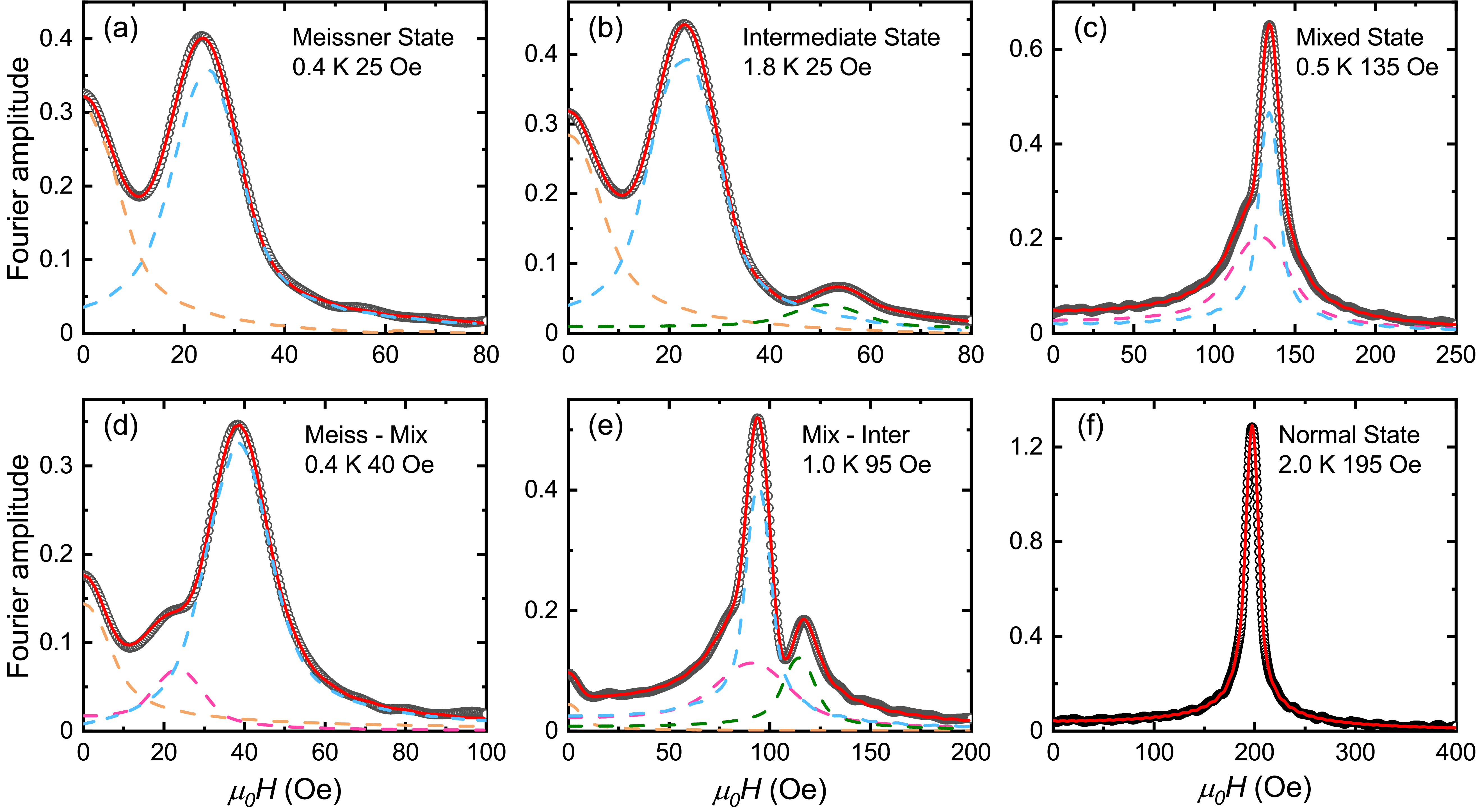}
		\caption{Fourier transformation of the total TF-\musr~spectra, showing the field distribution of the local field probed by muons. The figure illustrates the typical signal observed in the (a) Meissner, (b) Intermediate, (c) Mixed, (d) Meissner-mixed, (e) Mixed-intermediate, and (f) Normal states. The black circles and the red solid lines represent the Fourier amplitude for the raw TF-\musr~data and the fitting curves, respectively. The dashed lines represent the contribution of different magnetic signals to the total Fourier amplitude. The dashed lines in orange, green, and purple represent the zero-field Meissner signal, the intermediate state signal, and the mixed state signal, respectively. The blue dashed line represent the contribution of the sum of the normal state and the background signal.}
		\label{fig: TF2}
	\end{center}
\end{figure*}

For a Type-I/Type-II superconductor, the field distribution inside of the sample is complicate due to the possibles of Meissner, intermediate, IMS, mixed, and normal states. Therefore, Fourier transformations (FFTs) analysis was first applied. By analyzing the position and number of line shapes in FFTs curves, the state of the sample at different temperatures and fields can be preliminarily evaluated. Fig.~\ref{fig: TF2} shows the FFTs analysis of a set of raw TF-\musr~spectra, from which six different phases can be distinguished. 

Fig.~\ref{fig: TF2}(a)-(c) show the FFTs of the TF spectra, which are characteristic of the typical Meissner, intermediate, and mixed states, respectively. At $T=0.4$~K, $H=25$~Oe, which is well below the critical field at this temperature, \IrGa~is in the Meissner state. Besides the background signal, the FFTs result shows a strong component at zero magnetic field. The absence of any additional magnetic signals implies that the magnetic field is completely expelled from the sample. In our TF-\musr\ measurements, about 40\%~of muons stop in the silver sample holder and form the background signal. The background signal contributes a peak in the FFTs with a height of approximately 0.35-0.4, which exists in all our measurements. 

For a Type-I superconductor, intermediate state is induced by the demagnetization effect, which leads to a coexistence of magnetic fields with regions of zero field and internal field at $H_{c}$. Any real Type-I superconductor will have a nonzero demagnetizing factor $\eta$ determined only by the shape of the sample. When the applied field $H_a$ is in the range $(1-\eta)H_{c} < H_{a} < H_{c}$, the sample will enter into the intermediate state. At $T=1.8$~K, $H=25$~Oe, \IrGa~exhibits an intermediate-state behavior. As shown in Fig.~\ref{fig: TF2}(b), a peak at about 58~Oe, almost equivalent to the critical field at 1.8 K, is observed, as well as a peak at zero field. 

At $T=0.5$~K, $H=135$~Oe, \IrGa~is found in the mixed state. As a distinctive feature of Type-II superconductor, the mixed state is characterized by the quantized magnetic flux lattice. Consequently, there is a dispersion of internal fields, commencing from a minimum value and progressively intensifying until it reaches the field distribution's peak known as the saddle point. This saddle point denotes the most likely field value before declining with an extended tail, culminating in the maximum field value associated with the vicinity of the vortex core. This signal can be well described by a Gaussian distribution of the fields centered at the saddle point, which is around 125 Oe in this case (shown in Fig.~\ref{fig: TF2}(c)). Meanwhile, the absence of FFTs peak at zero field suggests that the full volume of the sample is in the mixed state. 

At $T=0.4$~K, $H=40$~Oe, as shown in Fig.~\ref{fig: TF2}(d), an unusual coexistence of the Meissner and the mixed state is observed. Neither Type-I nor Type-II superconductors would show this Meissner-mixed state. Such a state arises when vortices have weak attractive interaction. However, the most intriguing aspect is depicted in Fig.~\ref{fig: TF2}(e) and is referred to a so-called IMS. At $T=1.0$~K, $H=95$~Oe, peaks centered at the zero field and 120~Oe shows the behavior of the intermediate state. Meanwhile, a Gaussian distribution of the field centered at 90~Oe symbolizes the mixed state. This state signifies the simultaneous presence of both Type-I and Type-II superconductivity in the sample. It serves as an experimental proof of the novel coexistence of superconducting Type-I and Type-II \musr~responses. 

At $T=2.0$~K, $H=195$~Oe, \IrGa~turns into the normal state, and the field penetrates the entire volume of the sample. The FFTs shows only one peak centered at the applied field 195~Oe. 

\begin{figure*}[ht]
	\begin{center}
		\includegraphics[clip=,width=17cm]{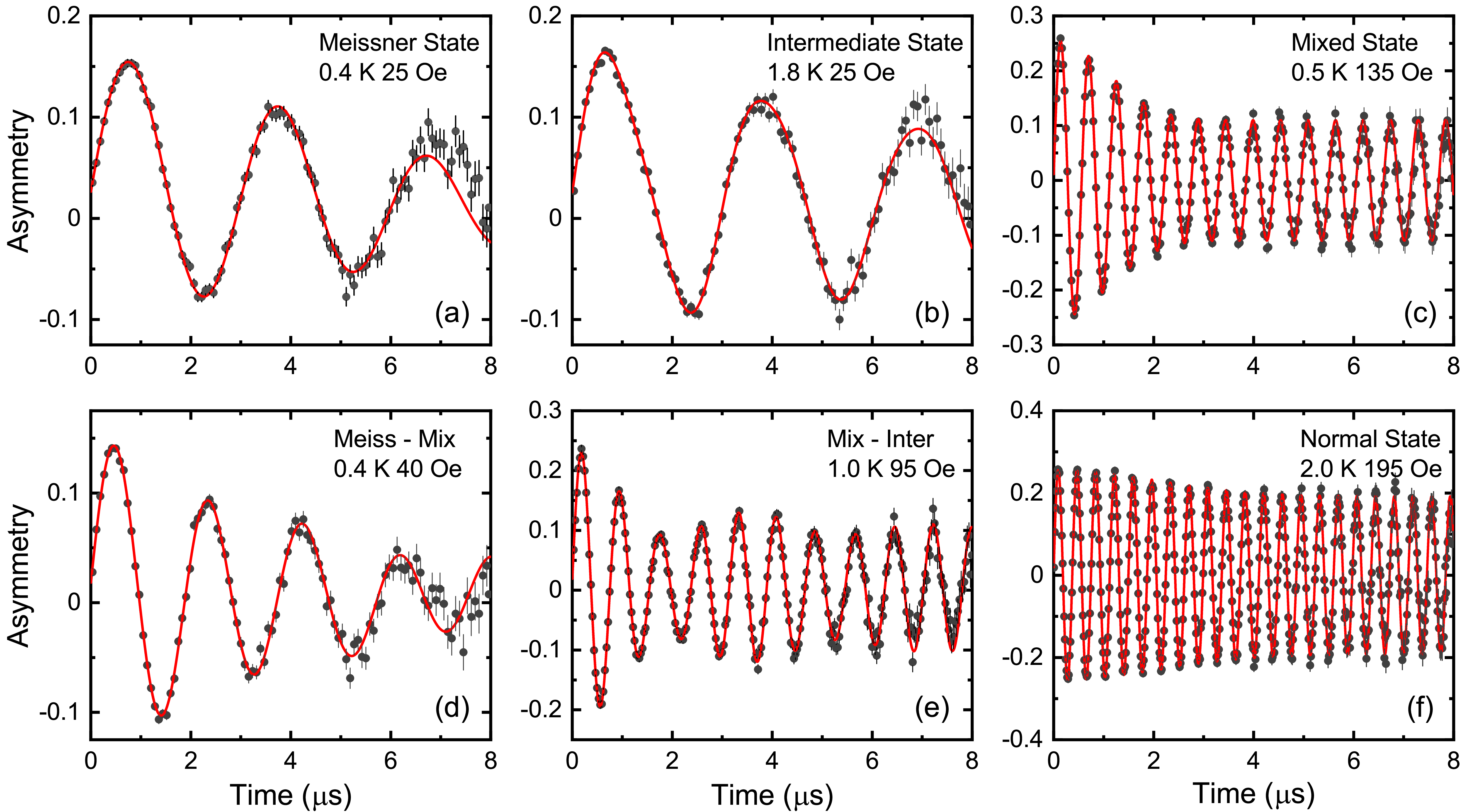}
		\caption{TF-\musr~time spectra collected at different temperatures and applied fields. The figure illustrates the typical signal observed in the (a) Meissner, (b) Intermediate, (c) Mixed, (d) Meissner-mixed, (e) Mixed-intermediate, and (f) Normal state. The solid curves are fits to the data using Eq.~(\ref{eq:TF}).}
		\label{fig: TF1}
	\end{center}
\end{figure*}

Having the information of the field distribution, the TF spectra can be analyzed by using the following formula:
\begin{equation}
	\begin{split}
		\label{eq:TF}
		{\rm Asy}(t) = A_{\rm{bg}}e^{-\lambda_{\rm{bg}}t}\cos(\omega_{\rm{bg}}t+\phi)\\
		+ A_{\rm{Meiss}}G_{\rm{ZF}}^{\rm{KT}}(\sigma_{\rm{KT}}, t) \\
		+
		\sum_{i=\rm{inter, mix, normal}}A_{i}\exp(-\frac{1}{2}\sigma_{i}^{2}t^{2})\cos(\omega_{i}t+\phi),
	\end{split}
\end{equation}
where the first term is due to the muons stopped into the silver sample holder, and the second term is the Kubo-Toyabe function representing the response of the static nuclear moments in the Meissner state. The relaxation rate $\sigma_{\rm{KT}}$~is fixed at the same value as in our ZF-\musr~experiment. The last term in Eq.~(\ref{eq:TF})~ includes three Gaussian distributions of non-zero magnetic fields in the intermediate, mixed, and normal state. It should be mentioned that all five terms in Eq.~(\ref{eq:TF})~do not exist at the same time in the fitting of one TF-\musr~spectrum. For instance, if the sample is in the mixed state, there will be no Meissner and intermediate state contributions in its TF-\musr~spectrum. A set of TF-\musr~spectra corresponding to the FFTs images in Fig.~\ref{fig: TF2} are shown in Fig.~\ref{fig: TF1}. The different TF-$\mu$SR time spectra and the successful fitting using Eq.~(\ref{eq:TF}) with different line shape information confirm each state of the sample.

\subsection{ZF-\musr}

\begin{figure}[ht]
	\begin{center}
		\includegraphics[clip=,width=8.5cm]{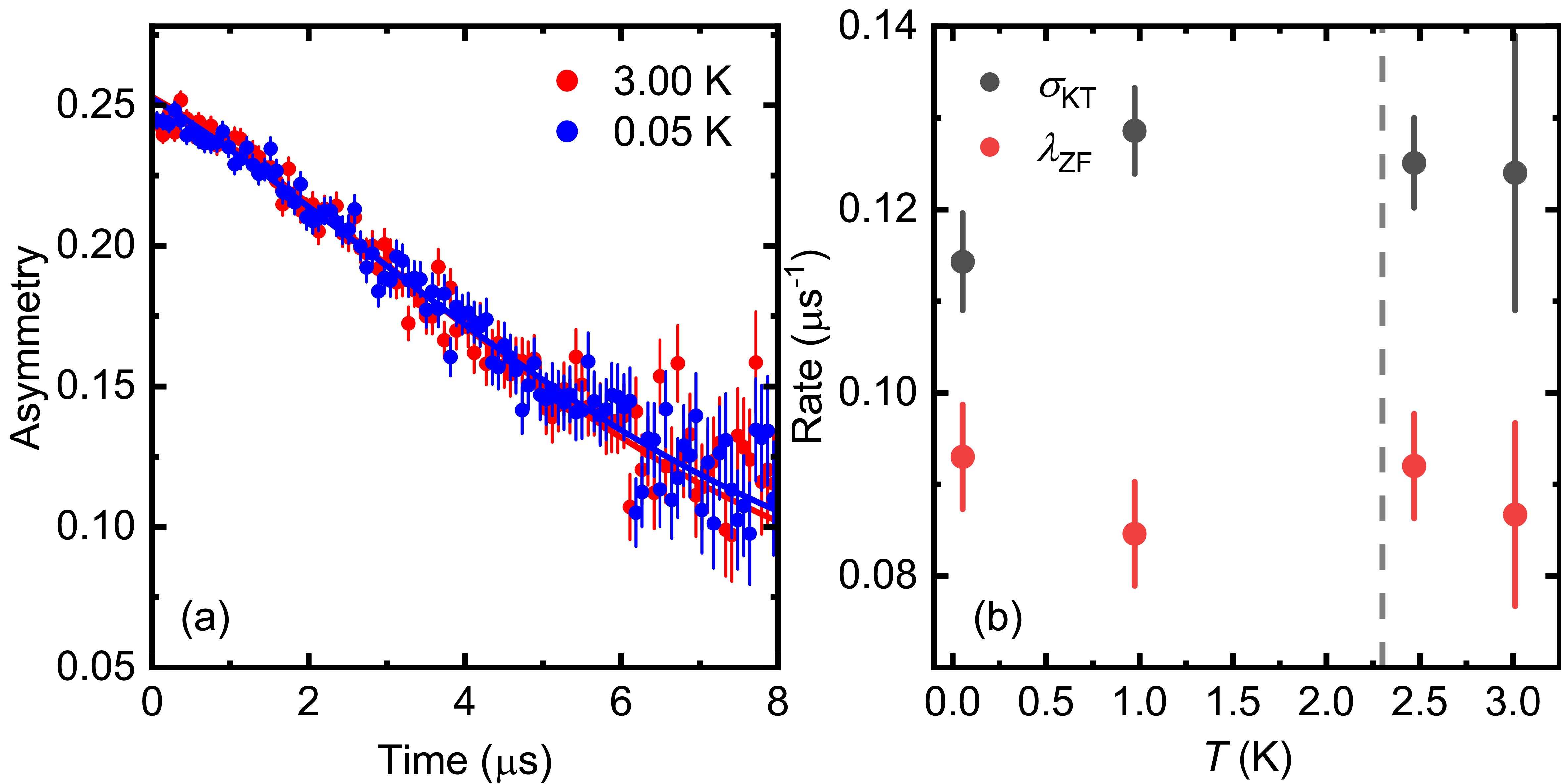}
		\caption{ZF-\musr. (a) \musr~ asymmetry spectra. Blue circles: superconducting state. Red circles: normal state. Solid curves: fits to the data with Eq.~(\ref{eq:ZF}). (b) Temperature dependences of the relaxation rates $\lambda_{\rm{ZF}}$\ and $\sigma_{\rm{KT}}$. Dashed line marks \Tc.}
		\label{fig: ZF}
	\end{center}
\end{figure}

ZF-\musr~spectra at representative temperatures are shown in Fig.~\ref{fig: ZF}(a). No significant difference can be observed between the data above and below \Tc~= 2.3 K. The \musr~asymmetry spectrum consists of two signal parts, which are from muons that stop in the sample and muons that stop in the silver sample holder, respectively. The spectra can be well fitted by the function:
\begin{equation}
	\label{eq:ZF}
	{\rm Asy}(t) = A_{0}[fe^{-\lambda_{\rm{ZF}}}G_{\rm{ZF}}^{\rm{KT}}(\sigma_{\rm{KT}}, t) + (1 - f)e^{-\lambda_{\rm{bg}}}]
\end{equation}
where $A_{0}$\ and $f$ represent the initial asymmetry and the fraction of muons stopping in the sample, respectively. The Kubo-Toyabe (KT) term~\cite{Kubo1979}
\begin{equation}
	\label{eq:KT}
	G_{\rm{ZF}}^{\rm{KT}}(\sigma_{\rm{KT}}, t) = \frac{1}{3}+\frac{2}{3}(1-\sigma_{\rm{KT}}^{2}t^{2})\exp(-\frac{1}{2}\sigma_{\rm{KT}}^{2}t^{2})
\end{equation}
describes a Gaussian distribution of randomly oriented static local fields with the distribution widths $\delta B_{\rm{G}}=\sigma_{\rm{KT}}/\gamma_{\mu}$, where $\sigma_{\rm{KT}}$\ is the relaxation rate.

The temperature dependences of the relaxation rate $\lambda_{\rm{ZF}}$\ and $\sigma_{\rm{KT}}$\ are shown in Fig.~\ref{fig: ZF}(b). No significant change crossing \Tc~is observed down to 0.05 K. This indicates that there is no spontaneous magnetic field appearing in the superconducting phase. This is consistent with the previous scanning SQUID result~\cite{Sumiyama2016}.

\section{Discussion}

\begin{figure}
	\begin{center}
		\includegraphics[clip=,width=8.5cm]{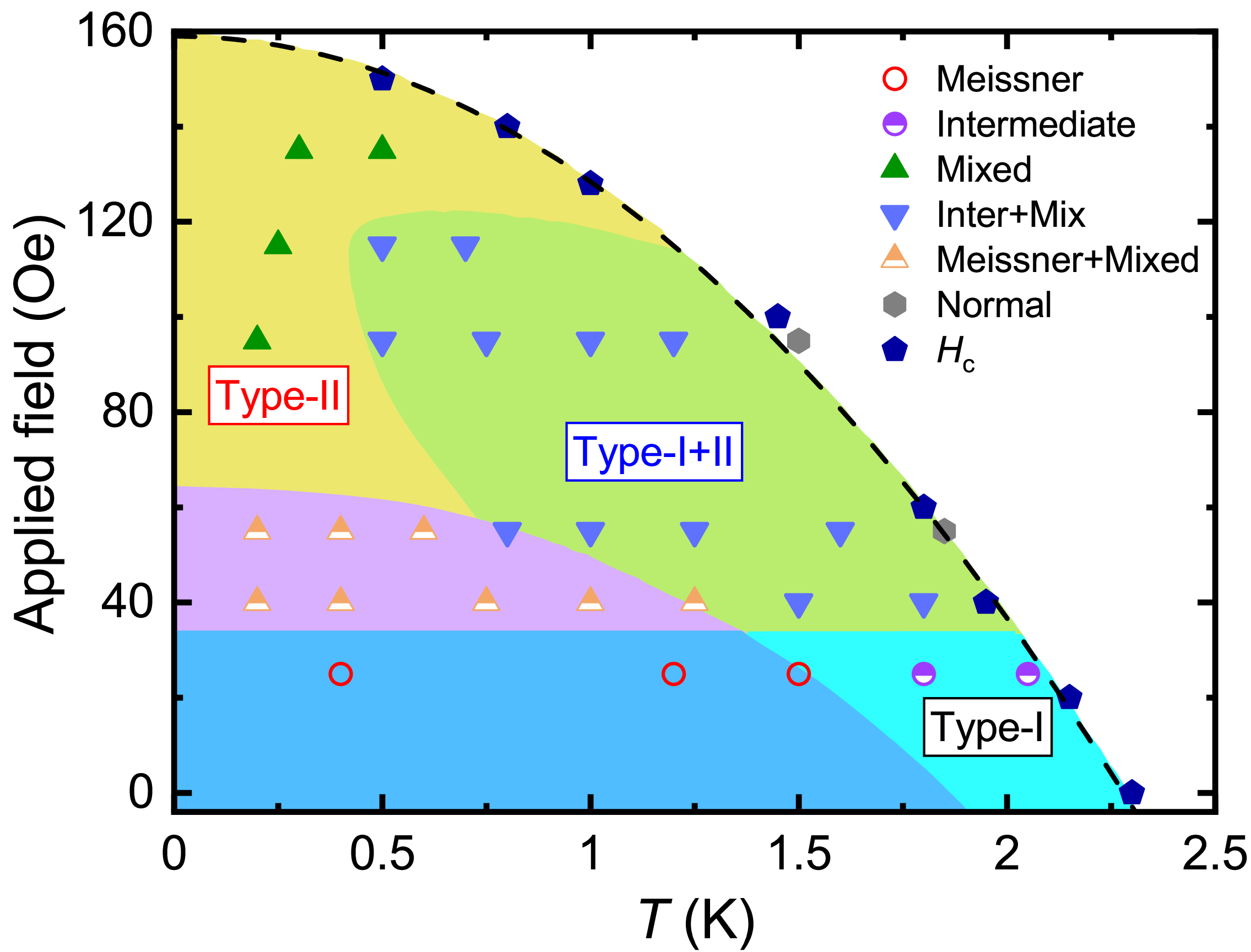}
		\caption{Different superconducting phases of \IrGa~as described in the text; the types refer to different \musr~responses shown in Fig.~\ref{fig: TF1} and Fig.~\ref{fig: TF2}. }
		\label{fig: PD}
	\end{center}
\end{figure}

TF-\musr~data reveal different states in \IrGa~at various fields and temperatures. These are summarized in a $H$-$T$ phase diagram in Fig.~\ref{fig: PD}.  

There are five different states in the superconducting state of \IrGa. In addition to the well-known Meissner, mixed and intermediated states, two unconventional states, Meissner-mixed and the IMS are observed. The coexistence of the vortex clusters and Meissner domains in \IrGa~is also observed in a relatively broad regime in the phase diagram. It should be mentioned that the critical magnetic field $H_c$ in the phase diagram comes from the measurement results of the specific heat. Due to the limitation of the MPMS temperature range, the difference between $H_c$ and $H_{c2}$ below 1.8 K cannot be obtained from the dc magnetization measurement. Judging from the isothermal magnetization curve at 1.8 K, the $H_c$ and $H_{c2}$ of the sample are very close, roughly 50 Oe and 60 Oe respectively, indicating that the GL parameters of the sample are very close to $\frac{1}{\sqrt{2}}$. Approximating the model by a single-component theory, the inhomogeneous vortex states in single-component superconductors can be induced by defects in a Type-II superconductor or by tiny attraction caused by various non-universal microscopic effects beyond the GL theory~\cite{Brandt1995, Babaev2010}. However, in \IrGa, the attractive interaction between vortices is observed over a broad regime in the phase diagram, which is inconsistent with the description of the single-component theory~\cite{Babaev2005}. 

Meanwhile, the phase diagram of \IrGa~is similar to that of the previous reported Type-I/Type-II superconductors ZrB$_{12}$~\cite{Biswas2020} and PdTe$_2$~\cite{Singh2019, Le2019}. However, the mechanism of the Type-I/Type-II superconductivity in ZrB$_{12}$ and PdTe$_2$ is different, which motivates us to discuss the Type-I/Type-II superconductivity from different scenarios.

\subsection{Multi-band superconductor?}

Considering the potential multi-gap behavior inferred from the low-temperature specific heat measurements, the Type-I/Type-II behavior of \IrGa~may be understood using the multi-band scenario. For a multi-band superconductor, if there are two superconducting coherence lengths, and the superconducting penetration depth is located between these two lengths, it may exhibit Type-I/Type-II behavior~\cite{Moshchalkov2009, Kogan2011, Babaev2011}. The coexistence of Type-I and Type-II signals can be understood as: one band exhibits Type-I properties, while the other band exhibits Type-II properties. This paradigm has been used to explain the Type-I/Type-II behavior of MgB$_2$ and ZrB$_{12}$~\cite{Nishio2010, Biswas2020}. In addition, there is a different theoretical explanation that the vortex attraction in the Type-I/Type-II behavior of multi-band superconductors may come from the interband proximity effect~\cite{Babaev2010}. In this case, only one band is superconducting while superfluid density is induced in another band via an interband proximity effect. For \IrGa, both the paradigms above cannot be ruled out. However, there is currently a lack of conclusive evidence that \IrGa~has multi-band features. Therefore, further research is needed to explain the Type-I/Type-II behavior of \IrGa~from a multi-band scenario.

\subsection{Topological surface state?}

The GL parameter $\kappa$ is defined as the ratio of the penetration depth $\lambda$ and the coherence length $\xi$ of superconductors. After considering the Pippard nonlocal electrodynamics, one can write $\kappa$ as a function of the electron mean free path $l$. The value of $\kappa$ is negatively related to $l$, and $l$ can be effected by certain kinds of topological surface effects, leading to the temperature and field dependence of $\kappa$. Such scenario is attributed to explain the Type-I/Type-II superconductivity in PdTe$_2$~\cite{Le2019, Singh2019}. Among NCS superconductors, \IrGa~is not the only sample that exhibits Type-I/Type-II superconductivity. Type-I and Type-II crossover at low temperatures has also been observed in NCS superconductors such as LaRhSi$_3$~\cite{Kimura2016} and NbGe$_2$~\cite{Lv2020}.  For these NCS superconductors showing Type-I/Type-II features, they have a common feature that there is a large surface critical field $H_{c3}$, which is considered to exist some kind of surface topological state~\cite{Wiesmann1977, Lv2020}. Therefore, the Type-I/Type-II superconductivity in \IrGa~may be explained by the inhomogeneous electron mean free path $l$ due to the potential surface topological state in NCS superconductors. However, for LaRhSi$_3$ and NbGe$_2$, their current $H$-$T$ phase diagrams are obtained by electrical resistance, magnetic susceptibility, and mechanical point-contact spectroscopy (MPCS) measurement~\cite{Kimura2016, Zhang2021}. In their phase diagrams, the boundaries between phases such as intermediate states, mixed states, and IMS are relatively indistinct, making it difficult to judge the similarities and differences between them and the $H$-$T$ phase diagram of \IrGa. More theoretical and experimental studies are required to explore the correlation between Type-I/Type-II superconductivity and NCS structure.

\section{Conclusion}

In summary, we have synthesized high-quality \IrGa~single crystals and examined their superconducting properties. We resolved the dispute about whether \IrGa~is Type-I or Type-II superconductor. Type-I/Type-II behavior of \IrGa~is observed from dc-magnetization measurement. The low-temperature electronic specific heat of \IrGa~exhibits the potential characteristics of multi-band superconductivity. The preservation of TRS is revealed by the ZF-\musr~ technique. TF-\musr~experiments were performed to map the phase diagram of \IrGa. The unconventional \musr~responses of the Meissner-mixed state and IMS, representing the Type-I/Type-II behavior, are shown in the phase diagram. Scenarios of the multi-band feature and the topological surface state scenario are discussed to explain the mechanism of the Type-I/Type-II superconductivity in \IrGa, while more careful studies are needed to obtain the accurate conclusion.

\begin{acknowledgments}

We are grateful for technical assistance from the TRIUMF Centre for Molecular and Materials Science. This research was funded by the National Natural Science Foundations of China, No.~12174065, the National Key Research and Development Program of China, No. 2022YFA1402203, and the Shanghai Municipal Science and Technology Major Project Grant, No.~2019SHZDZX01.  Research at CSU-Fresno was supported by NSF DMR-1905636, and work at CSU-LA was partially supported by NSF grant HRD-1826490.

\end{acknowledgments}

%
\end{document}